\documentclass[conference]{IEEEtran}
\usepackage{acronym}  % Or use \usepackage[acronym]{glossaries} if using the glossaries package
\usepackage{algpseudocode}
\usepackage{graphicx}
\usepackage{amsmath}
\usepackage{amssymb }
\usepackage{graphicx}
\usepackage[caption=false,font=small,labelfont=sf,textfont=sf]{subfig}
\usepackage{amsmath}
\usepackage{amssymb}
\usepackage{amsthm}
\usepackage{lipsum}
\usepackage{xcolor}
\usepackage[linesnumbered,ruled,vlined]{algorithm2e}
\usepackage[2]{editing}

\theoremstyle{definition}

\usepackage[ruled,vlined]{algorithm2e}
\def\BibTeX{{\rm B\kern-.05em{\sc i\kern-.025em b}\kern-.08em
    T\kern-.1667em\lower.7ex\hbox{E}\kern-.125emX}}

\usepackage[shortcuts,acronym]{glossaries}

\makeglossaries % generates the acronym list
\begin{document}

\title{Failure-Aware Access Point Selection for Resilient Cell-Free Massive MIMO Networks}
\author{ Mostafa~Rahmani~Ghourtani\textsuperscript{1},
Junbo Zhao\textsuperscript{1}, Yi Chu\textsuperscript{1}\\ 
Hamed Ahmadi\textsuperscript{1}, 
David Grace\textsuperscript{1},
Alister~G.~Burr\textsuperscript{1}
        \\
    \IEEEauthorblockA{\textsuperscript{1}School of Physics, Engineering and Technology, University of York,
     \\
    {\rmfamily\itshape \{rahmani.mostafa, Junbo.zhao, Yi.chu, hamed.ahmadi, david.grace, alister.burr\}@york.ac.uk}, 
    }

\thanks{The work presented in this paper was funded by the UK Department for Science, Innovation and Technology under project YO-RAN.}}

\markboth{\today}%
{Shell \MakeLowercase{\textit{et al.}}: Bare Demo of IEEEtran.cls for IEEE Journals}

\maketitle
\begingroup\renewcommand\thefootnote{}
\footnotetext{The work presented in this paper was funded by the UK Department for Science, Innovation and Technology under project YO-RAN.}
\endgroup

\begin{abstract}

This paper presents a Failure-Aware Access Point Selection (FAAS) method aimed at improving hardware resilience in cell-free massive MIMO (CF-mMIMO) networks. FAAS selects APs for each user by jointly considering channel strength and the failure probability of each AP. A tunable parameter  \(\alpha \in [0,1]\)  scales these failure probabilities to model different levels of network stress. We evaluate resilience using two key metrics: the minimum-user spectral efficiency, which captures worst-case user performance, and the outage probability, defined as the fraction of users left without any active APs. Simulation results show that FAAS maintains significantly better performance under failure conditions compared to failure-agnostic clustering. At high failure levels, FAAS reduces outage by over 85\% and improves worst-case user rates. These results confirm that FAAS is a practical and efficient solution for building more reliable CF-mMIMO networks.

%Resilience is assessed using minimum-user spectral efficiency and user outage probability. Simulations show that FAAS significantly outperforms failure-agnostic clustering, reducing outage by over 85\% at high failure levels and improving worst-case user performance. FAAS offers a practical and scalable solution for robust CF-mMIMO deployments.

%This paper introduces Failure-Aware AP Selection (FAAS) to bolster hardware resilience in cell‑free massive MIMO (CF‑mMIMO) systems. In ultra-dense deployments, even low AP failure rates—due to power loss, hardware degradation, or synchronization faults—can severely impair service continuity and spectral efficiency. FAAS combines channel quality with individual AP failure probabilities when forming user associations. We evaluate resilience using minimum user spectral efficiency as the worst-case performance metric. A tunable failure-intensity parameter $\alpha \in [0,1]$ uniformly scales AP failure rates, enabling simulation across a spectrum of stress scenarios. Extensive simulations benchmark FAAS against failure-agnostic baselines, demonstrating that it preserves fairness and worst-case rates far more effectively as $\alpha$ increases. These results highlight FAAS as a provably effective, low-complexity enhancement for fault-tolerant CF‑mMIMO networks. %Results show that FAAS consistently maintains stronger minimum spectral efficiency under increasing hardware unreliability, highlighting its value as a practical and scalable solution for resilient CF-mMIMO operation.

\end{abstract}

\begin{IEEEkeywords}
Access point selection, Cell-free massive MIMO, Resilience, Spectral efficiency
\end{IEEEkeywords}

\section{Introduction}
%\mosi{How to define Quantitative resilience evaluation metrics?}
%\mosi{Which  Signal processing algorithm could be defined for resilient communications}
%\mosi{MIMO architectures (cloud RAN, multi-cloud MIMO, massive MIMO, cell-free MIMO) for resilience; How?}
%Cell-free massive MIMO (CF-mMIMO) systems consist of a large number of distributed access points (APs) cooperatively serving users without cell boundaries. One crucial advantage of CF-mMIMO is its inherent potential for resilience due to spatial diversity. In this focused analysis, we explore \textbf{resilience with respect to hardware failures} of APs. Specifically, we consider scenarios where each AP has a non-zero probability of failure, and investigate how \textbf{AP selection strategies} can improve system robustness.
Cell-free massive multiple-input multiple-output (CF-mMIMO) has emerged as a leading architecture for beyond-5G/6G wireless networks, generalizing classical massive MIMO into a distributed, cell-less paradigm \cite{ngo2017cell, chu2025testbed, ahmadi2025towards}. By deploying a large number of distributed access points (APs) that jointly serve all users, CF-mMIMO leverages macro-diversity, mitigates inter-cell interference, and ensures uniformly high data rates \cite{interdonato2019scalability, ghourtani2025link}. Unlike traditional cellular systems, CF-mMIMO significantly reduces cell boundaries and associated edge effects, enabling consistent quality of service (QoS) and ultra-reliable links \cite{eskandari2025network}. At the same time, the large number of distributed APs supports efficient MU-MIMO transmission, allowing the system to exploit spatial multiplexing gains that boost spectral efficiency while maintaining uniform coverage and link reliability \cite{rahmani2022deep2, rahmani2022deep, mohammadzadeh2025pilot}.

Despite these touted reliability benefits, the resilience of CF-mMIMO networks in the face of hardware failures has received surprisingly limited attention, in contrast to recent works highlighting resilience-by-design as a crucial paradigm for ensuring robust 6G communication networks \cite{khaloopour2024resilience}.  In practice, however, AP hardware can malfunction or fail (e.g. due to power outages, equipment faults, or maintenance issues), which poses a serious challenge to any distributed antenna system. Conventional cellular networks suffer outages when a base station fails, but a distributed CF network could be more fault-tolerant by design. For example, recent architectural proposals like the “radio stripes” concept suggest that node failures can be tolerated via internal routing mechanisms, thereby improving network robustness. This assumption holds particularly in high-density deployments, where overlapping AP coverage ensures that the failure of a few nodes has only a marginal impact on system-wide performance due to the inherent spatial redundancy of CF-mMIMO \cite{ngo2024ultradense}.

While resilience in CF-mMIMO is often discussed qualitatively \cite{chowdhury2023resilient}, only a few works offer detailed analysis. Sadreddini et al. \cite{sadreddini2023distributed} use Markov models to show how limited fronthaul capacity and long routing paths can disconnect UEs or degrade SINR. Weinberger et al. \cite{weinberger2023ris} demonstrate that RIS can passively enhance resilience by providing alternative paths, even without optimized phase settings. Elkeshawy et al. \cite{elkeshawy2025robust} propose a data-driven activity detector at the central processing unit (CPU) that remains accurate under impairments, highlighting robustness against practical impairments. In addition, \cite{jiang2025nonlinear} addresses hardware nonlinearity by modeling PA distortions and optimizing user association and power control to mitigate their effects. Overall, there remains a significant theoretical gap in understanding how probabilistic AP failures influence CF-mMIMO performance and what can be done to design resilient cell-free networks. %Issues such as how much spectral efficiency degrades as more APs fail and how to proactively mitigate failure impacts are largely unexplored in current research. This gap has important system-level implications: without resilience, the promised “uniformly great service” of CF-mMIMO could be compromised by random hardware outages, especially in ultra-dense deployments where the probability of some AP failing is non-negligible.

The theoretical novelty of this work lies in integrating hardware failure resilience into CF-mMIMO for the first time in a systematic way. \textcolor{black}{Rather than deriving closed-form analytical expressions for performance under failures, which remain highly complex due to combinatorial failure patterns, we propose a tractable modeling framework that incorporates probabilistic AP failures into AP selection and evaluation. By explicitly defining failure-aware user--AP associations and resilience metrics, our work bridges the gap between purely qualitative discussions of resilience and quantitative system-level analysis.} We show that even under moderate AP failure rates, intelligent AP selection can preserve much of the system’s spectral efficiency, whereas traditional failure-agnostic approaches suffer more pronounced degradation. 
The proposed \textit{Failure-Aware AP Selection} (FAAS) methodology offers a blueprint for making CF architectures failure-aware: network controllers can use failure probabilities (obtained from hardware health monitoring or historical data) to optimize user-AP associations proactively. 

To validate these claims, we evaluate FAAS under various probabilistic failure scenarios and compare the resulting user rates and fairness against baseline schemes without failure awareness. The introduction of resilience into CF-mMIMO, as pursued in this work, opens a new research direction to ensure that the next generation of CF-mMIMO networks can deliver on their promise of ubiquitous, reliable connectivity even in the presence of inevitable hardware failures. In the following, we detail the system model and assumptions, then present the FAAS strategy and its theoretical performance analysis under AP failure conditions.

\section{Problem Statement}

\begin{figure}
    \centering
    \includegraphics[width=1\linewidth]{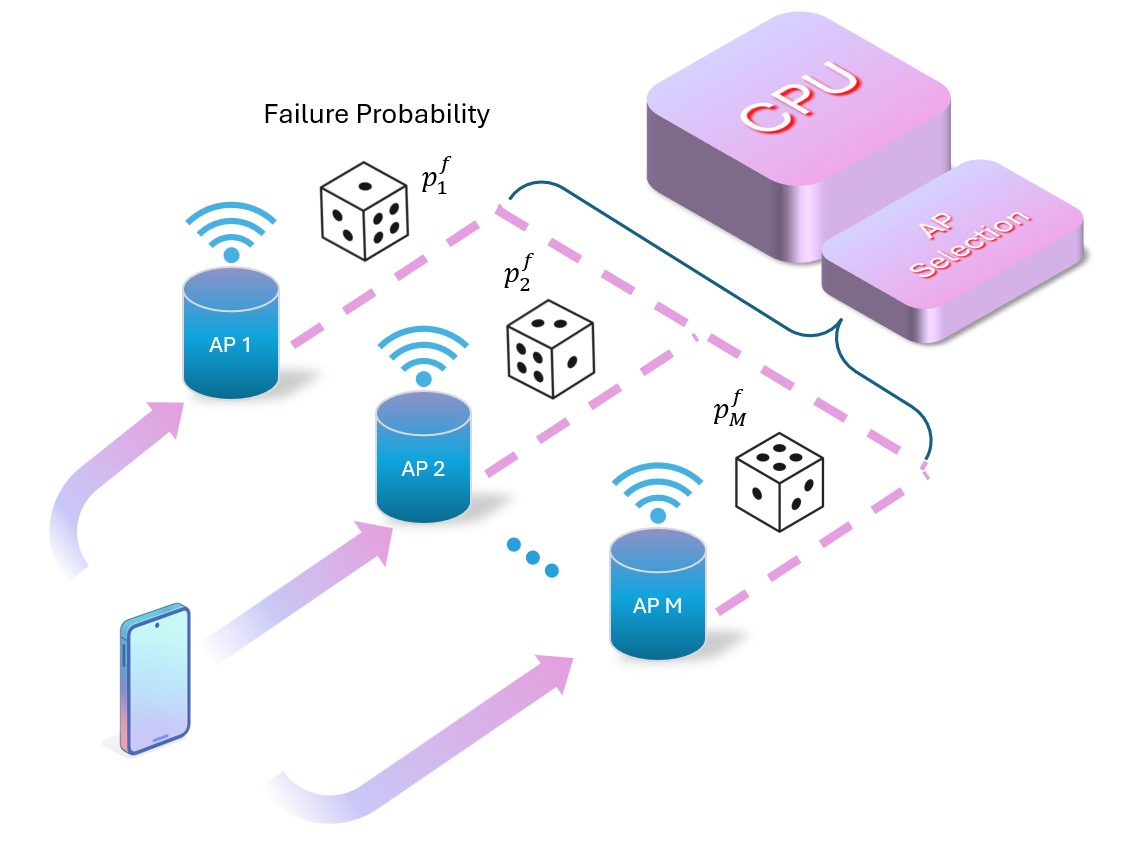}
    \caption{Failure-aware AP selection framework: each AP is associated with a failure probability $p_m^f$ and CPU selects APs based on both channel strength and failure probability.}
    \label{Arch}
    %\vspace{-0.5cm}
\end{figure}

As illustrated in Fig.~\ref{Arch}, APs in a CF-mMIMO network are susceptible to hardware failures caused by power loss, component degradation, or synchronization issues. If each AP fails independently with a probability $p_{m,0}^f$, the set of active APs becomes a random subset of the total $M$ deployed units. \textcolor{black}{In practice, $p_{m,0}^f$ can be estimated from hardware reliability statistics, field measurements, or predictive health monitoring of AP components. Typical values lie in the range $0.01$--$0.1$, corresponding to 1--10\% failure likelihoods as reported in radio access and power systems. In this work, we adopt such representative ranges to model realistic stress levels rather than targeting a specific failure mechanism, making the framework broadly applicable across different deployment scenarios. The assumption of independent failures provides a tractable and widely used baseline in reliability analysis; extending the framework to correlated failures (e.g., site-level or fronthaul outages) is an important open problem left for future work.}

%As illustrated in Fig.~\ref{Arch}, APs in a CF-mMIMO network are susceptible to hardware failures caused by power loss, component degradation, or synchronization issues. If each AP fails independently with a probability \(p_{m,0}^f\), the set of active APs becomes a random subset of the total \(M\) deployed units. \textcolor{black}{In practice, \(p_{m,0}^f\) can be estimated from hardware reliability statistics, field measurements, or predictive health monitoring of AP components. In this work, we adopt representative ranges of \(p_{m,0}^f\) to model realistic stress levels rather than targeting a specific failure mechanism, making the framework broadly applicable across different deployment scenarios.}

In such conditions, static user-to-AP associations can lead to service outages or severe performance degradation when assigned APs are unavailable. To address this challenge, we propose the FAAS scheme, which proactively integrates failure awareness into CF-mMIMO system design to enhance resilience. The key idea behind FAAS is to dynamically select a subset of APs for each user by accounting for the probability of AP failures. Unlike conventional schemes that assume a fixed, always-on set of APs, FAAS adapts the AP-user association strategy based on potential outages, leveraging the redundancy inherent in distributed AP deployments. Even if some APs randomly fail, the remaining active APs can maintain user service with minimal degradation. %Our simulation results highlights the limitations of failure-agnostic clustering, which can lead to sharp performance drops under adverse conditions, and demonstrates that FAAS significantly enhances robustness. By selecting reliable APs, FAAS enables fault-tolerant operation and improves overall network resilience in practical CF-mMIMO systems.

\section{System Model}
\subsection{Pilot Transmission and Channel Estimation}
We consider a CF-mMIMO system involving $M$ APs equipped with $N$ antennas and $K$ users uniformly distributed in the network. The signals from all APs are transmitted to the CPU via fronthauls and processed there. The flat-fading channel coefficient between the $m$-th AP and the $k$-th user is defined as $\textbf{g}_{mk} = \beta_{mk}^{1/2} \textbf{h}_{mk}$, where $\beta$ is the large-scale fading coefficient and $\textbf{h}_{mk}\sim\mathcal{CN}(\textbf{0},\textbf{R}_{mk})$ denotes small-scale Rayleigh fading with the spatial correlated matrix $\textbf{R}_{mk}\in\mathbb{C}^{N\times N}$.

To obtain the channel state information (CSI), we assume that $\tau_p$ mutually orthogonal pilot sequences, each of length $\tau_p$, are used. Let $\boldsymbol{\varphi}_k\in\mathbb{C}^{\tau_p\times1}$ with $||\boldsymbol{\varphi}_k||^2=\tau_p$ denote the pilot sequence assigned to the $k$-th user. The received pilot signal at the $m$-th AP is:
\begin{equation}
    \textbf{y}^\text{pilot}_m = \sum_{k=1}^{K}\sqrt{p_k}\textbf{g}_{mk}\boldsymbol{\varphi}_k^\text{T} + \textbf{z}^\text{pilot}_m,
\end{equation}
\noindent where $p_k$ represents the transmitted power for the $k$-th user, and $\textbf{z}^\text{pilot}_m\sim\mathcal{CN}(\textbf{0},\sigma_z^2\textbf{I}_N)$ denotes the noise vector at the $m$-th AP for the received pilot signal. Using the same approach \cite{bjornson2019making}, the coarse estimate is computed by $\check{\textbf{g}}_{mk} = \frac{1}{\sqrt{\tau_p}}\textbf{y}^\text{pilot}_m\boldsymbol{\varphi}_k^*$. Then, the MMSE estimate of $\textbf{g}_{mk}$ is:
\begin{equation}
    \hat{\textbf{g}}_{mk} = \sqrt{p_k\tau_p}\beta_{mk}\textbf{R}_{mk}\boldsymbol{\Psi}_{mk}^{-1}\check{\textbf{g}}_{mk},
\end{equation}
\noindent where 
\begin{equation}\label{Psi}
    \boldsymbol{\Psi}_{mk} = \sum_{k'\in S_k}\tau_pp_{k'}\beta_{mk'}\textbf{R}_{mk'} +\sigma_z^2\textbf{I}_N,
\end{equation}
\noindent where $S_k$ denotes the subset of users assigned the same pilot sequence $k$.% For later use, the covariance matrix of the channel estimation error $\tilde{\textbf{g}}_{mk}$ is defined by:

%\begin{equation}
%\begin{split}
%  &=  \mathbf{C}_{mk} \mathbb{E}\{\tilde{\textbf{g}}_{mk}\tilde{\textbf{g}}_{mk}^\text{H}\}  = \mathbb{E}\{(\textbf{g}_{mk}-\hat{\textbf{g}}_{mk})(\textbf{g}_{mk}^\text{H}-\hat{\textbf{g}}_{mk}^\text{H})\} \\
   % &= \mathbb{E}\{\textbf{g}_{mk}\textbf{g}_{mk}^\text{H}-\hat{\textbf{g}}_{mk}\hat{\textbf{g}}_{mk}^\text{H}\} 
  %  = \beta_{mk}\mathbf{R}_{mk} - \beta_{mk}^2\mathbf{R}_{mk}\boldsymbol{\Psi}_{mk}^{-1}\mathbf{R}_{mk} \\
%    &= (\mathbf{I}_N-\beta_{mk}\mathbf{R}_{mk}\boldsymbol{\Psi}_{mk}^{-1})\beta_{mk}\mathbf{R}_{mk}
%\end{split}
%\end{equation}

\subsection{Uplink Data Transmission}
For the uplink data transmission, the received signal at the $m$-th AP is:
\begin{equation}
    \textbf{y}_m = \sum_{k=1}^{K}\textbf{g}_{mk}x_k + \textbf{z}_m,
\end{equation}

\noindent where $x_k$ is the data transmitted by the $k$-th user with power $\sigma_x^2$, and $\textbf{z}_m\in\mathbb{C}^{N\times1}$ denotes the noise vector for the data transmission. At the CPU, signals from selected APs are combined by the weight $\textbf{w}_{mk}\in\mathbb{C}^{1\times N}$ to detect the data $x_k$ as follows:
\begin{equation}
\begin{split}
    \hat{x}_k &= \sum_{m=1}^{M}\textbf{w}_{mk}\textbf{D}_{mk}\textbf{y}_m \\
    & = \textbf{w}_{k}\textbf{D}_{k}\textbf{g}_kx_k + \sum_{k'\neq k}^{K}\textbf{w}_{k}\textbf{D}_{k}\textbf{g}_{k'}x_{k'} + \textbf{w}_{k}\textbf{D}_{k}\textbf{z},
\end{split}
\end{equation}
\noindent where $\textbf{w}_k = [\textbf{w}_{1k},\dots,\textbf{w}_{Mk}]$, $\textbf{g}_k=[\textbf{g}_{1k}^\text{T},\dots,\textbf{g}_{Mk}^\text{T}]^\text{T}$, and the noise $\textbf{z}=[\textbf{z}_1^\text{T},\dots,\textbf{z}_M^\text{T}]^\text{T}$. The selected APs to serve user $k$ is a subset $\mathcal{M}_k\subset\{1,\dots,M\}$, and it can be determined by a block-diagonal matrix $\textbf{D}_k=\text{diag}(\textbf{D}_{1k},\dots,\textbf{D}_{Mk})\in\mathbb{C}^{M\!N\times M\!N}$, where
\begin{equation}\label{Dmk}
  \textbf{D}_{mk} = 
  \begin{cases}
    \textbf{I}_{N},      & \text{if} \quad m\in\mathcal{M}_k \\
    \textbf{0}_{N},      & \text{if} \quad m\notin\mathcal{M}_k
  \end{cases}
\end{equation}

Then, the uplink spectral efficiency (SE) for the user $k$ can be expressed by:
\begin{equation}
    \text{SE}_k = \frac{\tau_u}{\tau_u+\tau_p}\mathbb{E}\big\{\text{log}_2(1+\text{SINR}_k)\big\},
\end{equation}
\noindent where $\tau_u$ is the length of data for uplink transmission within one coherence interval, and $\text{SINR}_k$ denotes the signal-to-interference-plus-noise ratio (SINR) for user $k$, which is given by (\ref{SINR}) as stated in \cite{bjornson2020scalable}.

\begin{equation}\label{SINR}
    \text{SINR}_k = \frac{p_k|\textbf{w}_k\textbf{D}_k\hat{\textbf{g}}_k|^2}{\sum_{k'\neq k}^{K}p_{k'}|\textbf{w}_{k'}\textbf{D}_{k'}\hat{\textbf{g}}_{k'}|^2+\textbf{w}_k\boldsymbol{\zeta}_k\textbf{w}_k^\text{H}},
\end{equation}
\noindent where $\boldsymbol{\zeta}_k=\textbf{D}_k(\sum_{k'=1}^{K}p_{k'}\textbf{C}_{k'}+\sigma_z^2\textbf{I}_{M\!N})\textbf{D}_k$, and $\textbf{C}_k=\text{diag}(\textbf{C}_{1k},\dots,\textbf{C}_{Mk})$. 

%\section{AP Selection for Resilient}
\subsection{User-centric Network with Dynamic Cooperative Cluster}
To illustrate the failure unaware AP selection, we reference the scalable user-centric CF-mMIMO described in \cite{bjornson2020scalable} as a benchmark. In \cite{bjornson2020scalable}, the authors adopted dynamic cooperative clustering (DCC) to select the AP cluster that serves a specific user. This approach can be summarized as follows: the user first selects its Master AP based on the strongest large-scale fading coefficient; the pilot with the least pilot contamination (dominated by (\ref{Psi})) observed by the Master AP is then assigned to the user; finally, given a threshold, if the neighboring AP has a slightly lower channel gain compared to the Master AP, the AP cluster is selected.

We assume that only a small subset of users contribute to the main interference, as partial APs serving the current user may also serve other users. Therefore, partial MMSE (P-MMSE) is used to maximize the SINR in (\ref{SINR}) for user $k$, as demonstrated in \cite{bjornson2020scalable}, and is given by:

\begin{equation}
\begin{split}
    \textbf{w}_k &= p_k\hat{\textbf{g}}_k^\text{H}\textbf{D}_k\Big[\sum_{k'\in\mathcal{P}_k}p_{k'}\textbf{D}_k\hat{\textbf{g}}_{k'}\hat{\textbf{g}}_{k'}^\text{H}\textbf{D}_k \\
 &+\textbf{D}_k\bigg(\sum_{k'\in\mathcal{P}_k}p_{k'}\textbf{C}_{k'}+\sigma_\text{z}^2\textbf{I}_{M\!N}\bigg)\textbf{D}_k\Bigg]^{\dagger},
\end{split}
\end{equation}

\noindent where $\dagger$ denotes the Moore–Penrose pseudo-inverse, and the index $k'$ is included in the set $\mathcal{P}_k=\{k':\textbf{D}_k\textbf{D}_{k'}\neq\textbf{0}_{M\!N}\}$, if user $k'$ is partially served by the same APs that serve user $k$.

\section{Failure-Aware Access Point Selection}

This section presents the proposed FAAS mechanism. We equip the system with a tunable reliability model, define a combined utility metric for AP selection, and formulate both average and worst-case spectral efficiency under probabilistic failure conditions.

\subsection{FAAS Algorithm with Stress-Level Parameter}
\label{sec:faas_algorithm}

We introduce a \emph{failure intensity} parameter \(\alpha \in [0,1]\) to scale each AP’s baseline failure probability \(p_{m,0}^f\), reflecting conditions such as hardware faults, power outages, or fronthaul disruptions:
\begin{equation}
    p_m^f = \alpha \, p_{m,0}^f.
\end{equation}
Here, \(\alpha = 0\) corresponds to a failure-free network, while \(\alpha = 1\)represents networks operating in challenging environments under maximum stress conditions. In the failure-aware scheme, the CPU is assumed to know the failure probabilities \(p_m^f\) from monitoring or predictive data. It constructs the serving AP set \(\mathcal{M}_k\) for each UE \(k\) to ensure reliable service by prioritizing APs with strong and reliable channels. The selection rule is given by:
\begin{equation}
\mathcal{M}_k = \left\{ m \,\middle|\, \frac{\sum_{m=1}^{\tilde{N}} \tilde{\beta}_{mk}(1 - p_m^f)}{\sum_{m'=1}^{N} \beta_{m'k}(1 - p_{m'}^f)} \geq \varepsilon \right\},
\end{equation}
where $\tilde{\beta}_{mk}$ are large-scale fading coefficients sorted in descending order of associated reliability, and $\tilde{N}$ is the smallest number of APs whose cumulative reliability-weighted gain meets the predefined threshold $\varepsilon$. The resulting set $\mathcal{M}_k$ defines the structure of the combining matrix $\mathbf{D}_k$, activating only the selected APs that contribute most significantly to reliable communication. \textcolor{black}{By construction, this reliability-weighted selection directly reduces the probability of user outage, since APs with higher survival probability $(1-p_m^f)$ are favored, ensuring that the likelihood of all assigned APs failing simultaneously is minimized. To avoid fragile assignments where only a single AP is selected (i.e., $\tilde{N}=1$), which would undermine resilience if that AP fails, we impose a minimum cooperative cluster size of at least two APs per user. This guarantees that each user retains connectivity under a single-point failure. More generally, this floor on cluster size can be tuned depending on service requirements; for example, mission-critical deployments may enforce $\tilde{N}\geq 3$ for stronger redundancy.}

The threshold $\varepsilon \in (0,1)$ plays a key role in balancing reliability and cooperation size: it governs how many of the strongest and most reliable APs are selected to serve each user. In our simulations, $\varepsilon$ is fixed to a representative value (e.g., 0.9) to ensure sufficient robustness while maintaining sparsity in the user--AP association. \textcolor{black}{To ensure fairness, this thresholding and minimum cluster-size rule are applied consistently across both the proposed FAAS and the failure-agnostic clustering method, so that differences in performance stem solely from failure awareness rather than cluster size.}

%where $\tilde{\beta}_{mk}$ are large-scale fading coefficients sorted in descending order of associated reliability, and $\tilde{N}$ is the smallest number of APs whose cumulative reliability-weighted gain meets the predefined threshold $\varepsilon$. The resulting set $\mathcal{M}_k$ defines the structure of the combining matrix $\mathbf{D}_k$, activating only the selected APs that contribute most significantly to reliable communication. The threshold $\varepsilon \in (0,1)$ plays a key role in balancing reliability and cooperation size: it governs how many of the strongest and most reliable APs are selected to serve each user. In our simulations, $\varepsilon$ is fixed to a representative value (e.g., 0.9) to ensure sufficient robustness while maintaining sparsity in the user-AP association. Importantly, to enable fair benchmarking, we apply the same $\varepsilon$ value in both the proposed and the failure-agnostic clustering method, ensuring that differences in performance are solely due to failure awareness rather than selection size.

%where \(\tilde{\beta}_{mk}\) are large-scale fading coefficients sorted in descending order of associated reliability, \(\tilde{N}\) is the smallest number of APs meeting the predefined threshold \(\varepsilon\). The resulting set \(\mathcal{M}_k\) defines the structure of the combining matrix \(\mathbf{D}_k\), activating only selected APs. 

\subsection{Analytical Modeling of Failure Impact}
\label{sec:faas_modeling}

In our FAAS framework, AP failures are modeled as independent Bernoulli events. Specifically, each AP in $ \mathcal{M}_k $ remains active with probability $q = 1 - p_m^f$. Assuming AP failures are independent, the set of active APs for user~$k$ becomes $\textbf{D}_k$ , where each $m \in \mathcal{M}_k$ fails with probability $p_f$. The effective SINR is given by equation (\ref{SINR}).
Although AP failures introduce combinatorial randomness in which subsets of APs are operational, this can be abstracted using approximate techniques.  Treating each AP as a node that is independently “thinned” (retained) with probability $q$ leads to a valuable statistical approximation: the effective cluster size is reduced in expectation by this factor. While we omit detailed formulas here, analytic tools like \emph{binomial thinning}  \cite{kella2023binomial}, commonly used in point-process analysis of wireless networks, can derive tractable estimates for expectation and variance of SINR under failures.  %\textcolor{black}{This subsection therefore does not aim to provide exact closed-form results, but rather establishes a probabilistic modeling framework that captures how random AP failures propagate into effective user SINR and spectral efficiency. This framework underpins the resilience metrics introduced in Section IV-C and guides the design of the FAAS algorithm.}

%\paragraph{Qualitative Impact on SINR and SE}  
%Under thinning, the beamforming power scales approximately with the number of active APs, i.e. by $q\cdot |\mathcal{M}_k|$, while interference and noise similarly diminish. Hence, the effective SINR degrades roughly multiplicatively by a factor close to $q$. This gradual reduction means the resulting spectral efficiency is well-approximated by

%\paragraph{Minimum-user SE under Random Thinning}  
%Since the “worst-off” user often corresponds to one with a low nominal SINR, the thinning model allows us to express
\subsection{Resilience Metric}
\label{faas_resilience}

To evaluate the fault tolerance of our FAAS scheme, we consider two complementary resilience metrics that jointly capture service reliability and quality under AP failure. First, we use the minimum-user spectral efficiency: 
\begin{equation}
    \text{SE}_{\min} = \min_{k \in \{1,\dots,K\}} \mathbb{E}_{\mathcal{F}}[\text{SE}_k],
\end{equation} where the expectation is taken over all failure events \(\mathcal{F}\). This metric directly measures the worst-case user throughput, ensuring that our design does not marginalize cell-edge users or those served by less reliable APs. Alongside this, we monitor the average spectral efficiency:
\begin{equation}
    \overline{\text{SE}} = \frac1K \sum_{k=1}^K \mathbb{E}_{\mathcal{F}}[\text{SE}_k],
\end{equation}
but we emphasize \(\text{SE}_{\min}\) to ensure FAAS genuinely enhances resilience rather than merely elevating aggregate performance. \textcolor{black}{While our framework evaluates spectral efficiency in expectation over failure events, alternative reliability metrics based on the distribution of user rates (e.g., the 1\% outage rate or 99\% quantile SE) are indeed relevant. We adopt expectation-based metrics for analytical tractability and comparability with existing CF-mMIMO works \cite{bjornson2020scalable}, but acknowledge that quantile-based metrics could provide a finer-grained view of resilience. Extending FAAS to explicitly optimize such distributional guarantees is an important direction for future work.}

\begin{figure*}[t!]
\centering
\subfloat[\footnotesize $L = 400$ APs and $N = 1$ antenna ]{
%\hspace{-.15cm}
\includegraphics[height=55mm]
{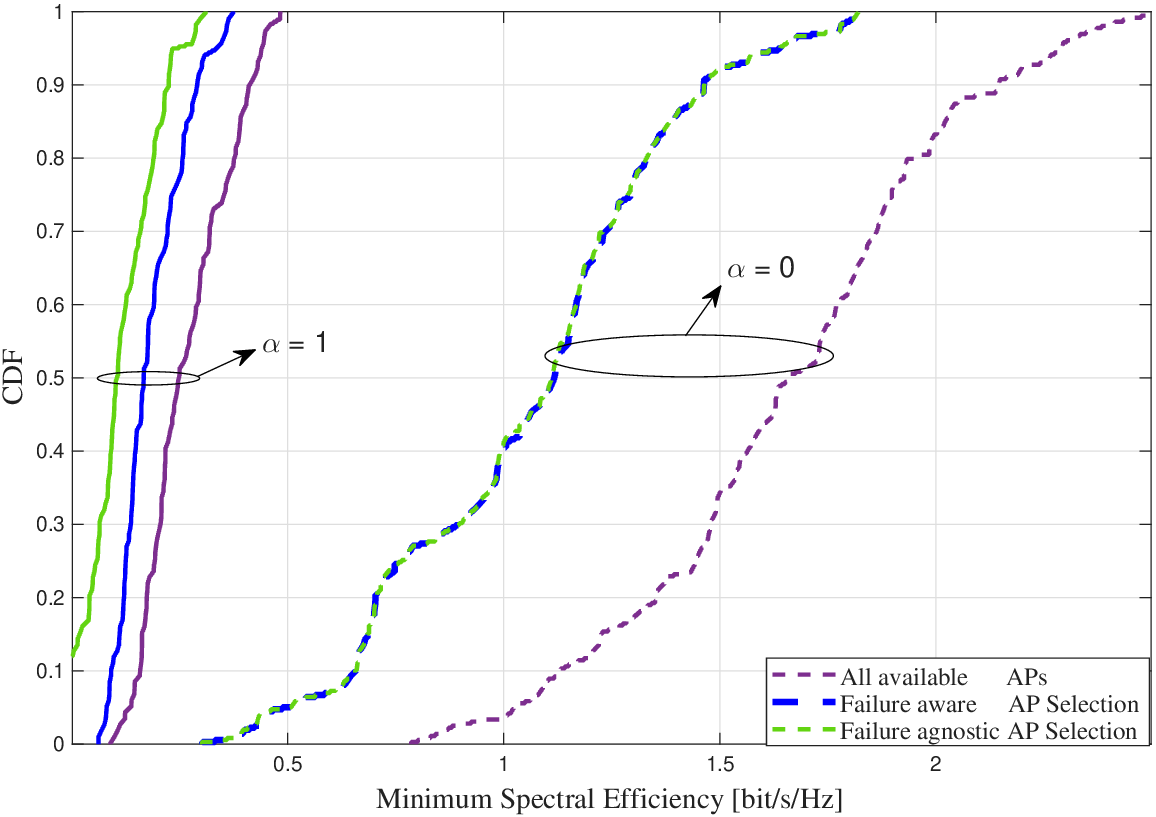}
\label{MinRate1}
}
\hspace{10pt} % 22
\vspace{-.02in}
\subfloat[ \footnotesize $L = 100$ APs and $N = 4$ antennas ]{
%\vspace{-.15cm}
\includegraphics[height=55mm]
{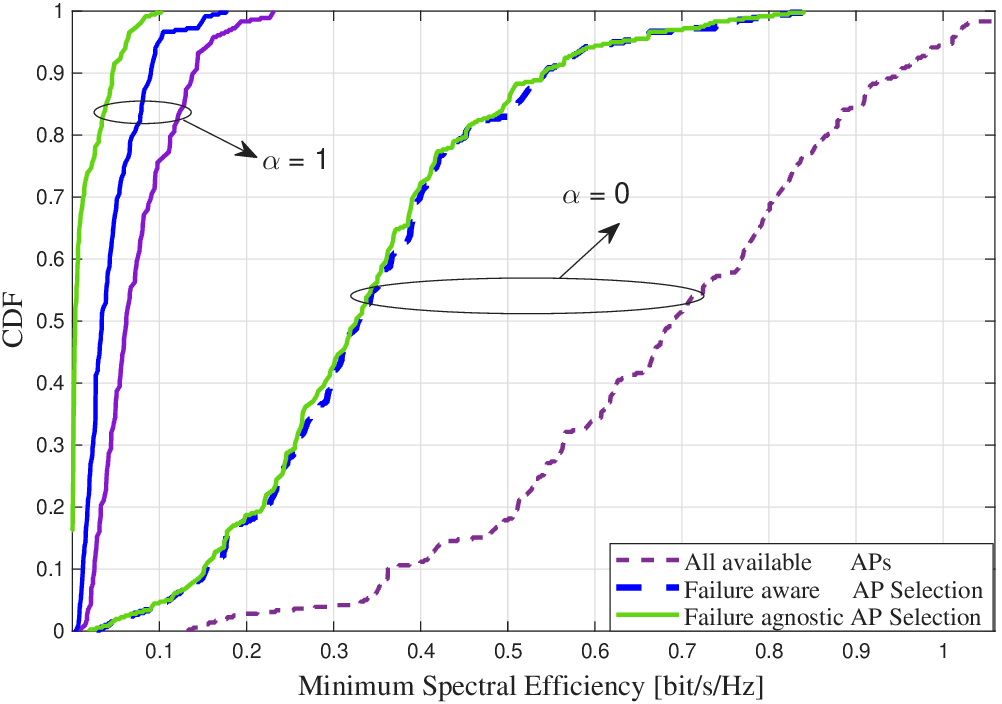}
\label{MinRate2}
}
\caption{CDF of uplink minimum user rate for cell-free
massive MIMO with MMSE combining and considering failure probability for APs, as previously mentioned \(\alpha = 0\) corresponds to a failure-free network, and \(\alpha = 1\) reflects maximum stress.}

%\vspace{-.55cm}
\label{MinRate}
\end{figure*}

%We define the \textbf{user outage probability} as the proportion of UEs that experience complete service disruption following AP failures, specifically when no active access points remain within their allocated cooperative cluster. Minimizing this outage rate is critical for network robustness, especially under moderate to high failure intensities, where maintaining at least basic connectivity offers far greater practical value than marginal gains in sum-rate. Collectively, these metrics provide a dual perspective on resilience: \(\text{SE}_{\min}\) ensures worst-user performance remains above acceptable levels, while the outage probability reflects the ability to sustain service at all. 

We define the \textbf{user outage probability} as the proportion of UEs that experience complete service disruption following AP failures, specifically when no active access points remain within their allocated cooperative cluster. \textcolor{black}{In the simplest case, if AP failures are independent and a user is served by $|\mathcal{M}_k|$ APs each with failure probability $p_m^f$, the outage probability reduces to $(p_m^f)^{|\mathcal{M}_k|}$. While this expression captures the intuition that outage is driven by the joint survival of all assigned APs, in our analysis we evaluate outage more generally by averaging over random failure events across the network. This provides a tractable metric even when failure probabilities differ among APs or when reliability-weighted clustering is used.}

\textcolor{black}{Importantly, the outage probability is not only an evaluation metric but also implicitly incorporated into the FAAS design: by weighting AP selection with their survival probability $(1-p_m^f)$ in (11) and by enforcing a minimum cooperative cluster size, FAAS proactively reduces the likelihood of user outage. Thus, FAAS both minimizes outage risk in the design phase and verifies the improvement through the defined metric.}

\medskip
\noindent\textbf{Connection to Failure‑Agnostic Baselines:}  
In traditional schemes where AP selection ignores failure risks, actual AP failures yield an unpredictable and often sharp drop in minimum-user SE, as active APs are removed arbitrarily. In contrast, FAAS anticipates this thinning by selecting more reliable APs, those with higher survival probability, even if their nominal channel is slightly weaker. This preemptive resilience is absent in failure-agnostic designs.
The combined utility metric makes FAAS \emph{stress-aware}, dynamically adapting selections as failure conditions worsen. The minimum-user SE metric enforces user fairness and robustness. 
%\textcolor{red}{Junbo: I believe you would like to define a factor $\alpha$ that corresponds to the failure probability $p^f$. This factor could be defined in this subsection, and it would be beneficial to relate it to equation (\ref{Dmk}) and the subset $\mathcal{M}_k$.}

\noindent\textbf{Network Topology Considerations:}
The proposed algorithms in this work have been described under the assumption of a star-topology architecture, in which all APs are connected to a CPU. This abstraction simplifies analysis and reflects a commonly adopted model in the literature. \textcolor{black}{The FAAS framework, however, is largely agnostic to the physical topology because the AP selection depends only on large-scale fading coefficients and failure probabilities, both of which can be obtained locally and aggregated by either a centralized CPU or distributed controllers. The user-centric nature of clustering ensures that only a small subset of neighboring APs cooperate for each user, which does not rely on full network centralization. Therefore, the same selection rule can be applied in star, hierarchical, or O-RAN style disaggregated architectures, as long as basic channel and reliability information is available.}

\section{Numerical Analysis}

To evaluate the performance of the proposed FAAS scheme, we consider a simulation setup similar to that in \cite{bjornson2020scalable}. Specifically, we simulate a CF-mMIMO network in which $K = 100$ UEs and $M$ APs are independently and uniformly distributed over a $2 \times 2$~km area. Two configurations are examined: (i) $M = 400$ single-antenna APs ($N = 1$), and (ii) $M = 100$ APs each equipped with $N = 4$ antennas. Both configurations yield the same spatial antenna density of 100 antennas/km$^2$, while maintaining a consistent user density of 25 UEs/km$^2$.

To mitigate edge effects and emulate an unbounded network, the wrap-around technique is applied, allowing accurate modeling of both path loss and interference. The simulation adopts the propagation model described in \cite{bjornson2020scalable}, which incorporates spatially correlated Rayleigh fading. All APs are positioned 10 meters above the UEs to enforce a realistic minimum path loss, reflecting practical urban deployment conditions.

We assume a coherence block length of $\tau_c = 200$ symbols, with $\tau_p = 10$ symbols allocated to orthogonal pilot transmission. The remaining $\tau_u = 190$ symbols are reserved for uplink data transmission. Each UE transmits with a fixed power of $p_k = 100$~mW, and the total system bandwidth is set to 20~MHz. Unless otherwise stated, all performance results correspond to the uplink phase, where both spectral efficiency and user outage probability are evaluated under varying levels of AP failure intensity $\alpha$. \textcolor{black}{For the baseline failure probability $p_{m,0}^f$, we consider values in the range $0.01$--$0.1$, consistent with reported reliability levels of commercial radio units and power systems. The stress parameter $\alpha \in [0,1]$ then scales these baseline values to emulate different network conditions, from nominal operation to highly stressed environments.}

\textcolor{black}{To prevent degenerate cases where a user would be associated with only a single AP, we enforce a minimum cooperative cluster size of two APs per user in all simulations. This practical safeguard reflects realistic deployment considerations and ensures that the resilience evaluation is not biased by fragile single-AP assignments.} We compare the proposed FAAS scheme against a baseline failure-agnostic clustering approach.

\begin{figure*}[t!]
\centering
\subfloat[\footnotesize $L = 400$ APs and $N = 1$ antenna ]{
%\hspace{-.15cm}
\includegraphics[height=56mm]
{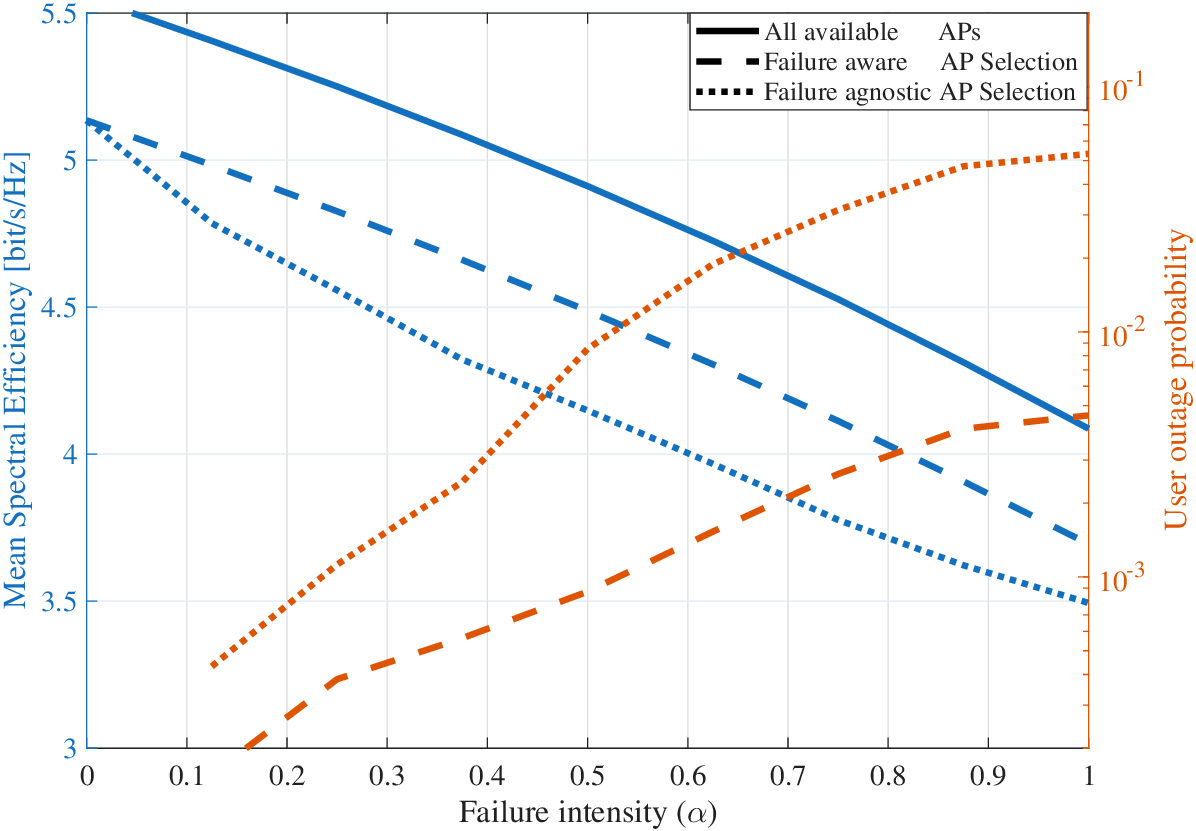}
\label{Mean1}
}
\hspace{10pt} % 22
%\vspace{-.02in}
\subfloat[ \footnotesize $L = 100$ APs and $N = 4$ antennas ]{
%\vspace{-.15cm}
\includegraphics[height=56mm]
{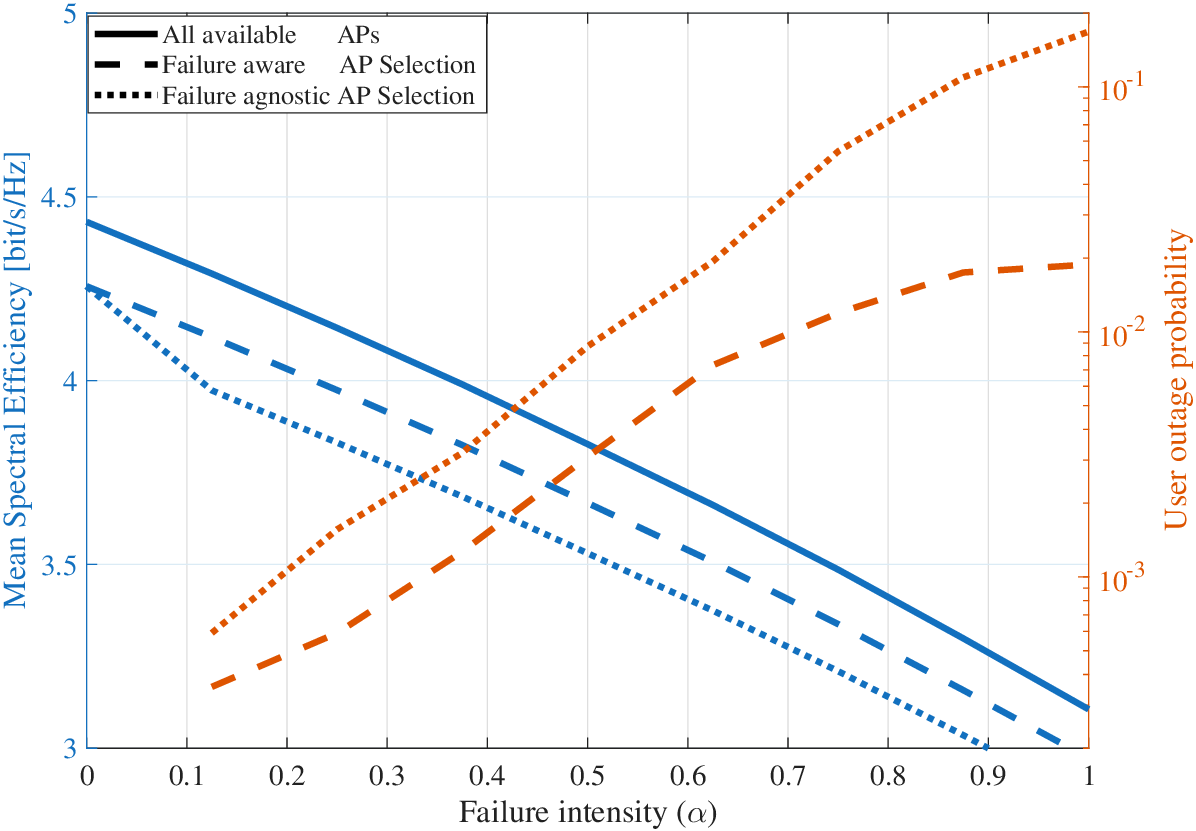}
\label{Mean2}
}
\caption{Impact of failure intensity $\alpha$ on mean spectral efficiency and user outage probability for CF-mMIMO with MMSE combining under different AP selection strategies. \textit{Note: The outage curve for ``All available APs'' is omitted as it remains zero for all $\alpha$; with a logarithmic scale, it is not visible.}}

%\vspace{-.55cm}
\label{Mean}
\end{figure*}

Fig.~\ref{MinRate} shows the CDF of the minimum uplink spectral efficiency for two CF-mMIMO setups and three AP selection schemes: All APs, Failure-Agnostic, and FAAS. When $\alpha = 0$, Failure-Agnostic and Failure-Aware perform identically since all APs are functional. The All APs case performs best across all settings but is unrealistic and unscalable. %At $\alpha = 1$, FAAS clearly outperforms the Failure-Agnostic method by prioritizing APs with higher reliability, thereby improving performance in the lower SE tail.

When $\alpha = 1$, failures occur according to the predefined AP failure probability. In this condition, the ``All available APs" case achieves the best performance, making it an idealized but unscalable benchmark. The failure-agnostic scheme suffers a noticeable performance degradation, particularly in the lower tail of the CDF, due to random AP selection that does not account for failure likelihoods. In contrast, the proposed FAAS approach significantly outperforms the failure-agnostic strategy by intelligently selecting APs based on their reliability. This confirms the benefit of incorporating failure awareness into AP selection for improving user fairness and network robustness under realistic failure conditions.

%Fig.~\ref{Mean} illustrates the average uplink spectral efficiency versus failure intensity $\alpha$ for two system configurations. As $\alpha$ increases, performance degrades due to a higher probability of AP outages. When $\alpha = 0$, the Failure-Aware and Failure-Agnostic schemes yield identical results, since no AP failures occur and reliability-aware selection offers no advantage. The All APs case achieves the highest spectral efficiency across all values of $\alpha$, but it is not scalable as it assumes full AP cooperation regardless of failures. As failures increase, the proposed Failure-Aware approach consistently outperforms the Failure-Agnostic baseline by selecting APs based on their availability probabilities, thus offering improved robustness under stressed conditions.
Fig.~\ref{Mean} presents the average spectral efficiency and user outage probability as a function of $\alpha$. As failure intensity increases, all schemes experience degradation. While the All APs scheme maintains the highest average SE, FAAS consistently outperforms Failure-agnostic clustering by selecting more reliable APs. This highlights FAAS's advantage in balancing performance and resilience.

Fig.~\ref{Mean} shows the effect of increasing failure intensity \(\alpha\) on both mean spectral efficiency and user outage probability for two CF-mMIMO configurations. As failures become more frequent, all schemes experience performance degradation. The \textit{All APs} strategy maintains the highest spectral efficiency and zero outage by assuming ideal full connectivity, though it is not scalable in practice. In contrast, the proposed FAAS approach consistently outperforms the failure-agnostic method by selecting APs based on their reliability, thereby enhancing both spectral efficiency and resilience.

FAAS significantly reduces the user outage probability, particularly under high \(\alpha\), where failure-agnostic clustering suffers from increased service disruption due to random AP assignment. By proactively incorporating failure awareness into AP selection, FAAS ensures stronger user connectivity even under stressed network conditions.

When comparing the two setups, many single-antenna APs (\(L = 400\), \(N = 1\)) versus fewer multi-antenna APs (\(L = 100\), \(N = 4\)), the single-antenna configuration demonstrates superior resilience. It achieves lower outage probabilities and higher average spectral efficiency across the range of \(\alpha\). This improvement stems from enhanced macro-diversity: UEs with weak links benefit more from the broader spatial distribution of single-antenna APs than from the local interference suppression offered by fewer, more capable APs.

\section{Conclusion}
This paper proposed a failure-aware AP selection strategy for CF-mMIMO systems, aiming to enhance network resilience under realistic hardware failure conditions. By integrating channel quality and AP-specific failure probabilities through a tunable stress parameter $\alpha$, the FAAS algorithm adaptively selects reliable APs per user. Simulation results demonstrate that FAAS significantly improves minimum spectral efficiency and reduces outage probability compared to failure-agnostic clustering, especially under moderate to high failure stress. Additionally, we observed that dense deployments of single-antenna APs offer better resilience than fewer multi-antenna APs due to increased macro-diversity. These findings underscore the importance of incorporating failure resilience as a core design aspect in CF-mMIMO networks. Future work will extend FAAS to disaggregated architectures like O-RAN, incorporating distinct failure models for different network components.  \textcolor{black}{In future work, FAAS could also be evaluated and optimized with respect to quantile-based performance metrics (e.g., 99\% user rate guarantees), providing even stronger resilience assurances beyond expectation-based analysis.}

%CF-mMIMO systems are inherently more resilient to hardware failures due to their distributed nature. However, random failures still degrade performance. By employing \textbf{AP selection strategies} that consider both channel quality and AP reliability, systems can maintain high service availability and user fairness, even in the presence of failing APs. This makes AP selection a vital tool for designing resilient next-generation wireless networks.
\bibliographystyle{IEEEtran}
\bibliography{refs}

\end{document}